\documentclass[prd,aps,nofootinbib,floatfix,11pt]{revtex4}
\usepackage{amsmath,graphicx,epsfig,amssymb,dsfont,mathtools,}
\usepackage[usenames]{color}
\usepackage{ulem} 
\usepackage{bigstrut}
\usepackage{slashed}
\usepackage{multirow}
\usepackage{subfigure}

\allowdisplaybreaks

\begin{document}
  \title{Production of singly charm pentaquark from meson-baryon interaction in B decay}

\author{Na Li}
\email{TS22180005A31@cumt.edu.cn}
\affiliation{School of Materials Science and Physics, China University of Mining and Technology, Xuzhou 221000, China}

\author{Ye Xing}
\email{Corresponding author. xingye_guang@cumt.edu.cn}
\affiliation{School of Materials Science and Physics, China University of Mining and Technology, Xuzhou 221000, China}

\author{Ke-Fan Jiang}
\email{TS23180004A31@cumt.edu.cn}
\affiliation{School of Materials Science and Physics, China University of Mining and Technology, Xuzhou 221000, China}
\begin{abstract}
  In the paper, we discuss the possible interpretation of the $J^P=1/2^-$ singly charm pentaquark as hadronic molecules. With the effective Lagrangian method, we further analyze the production properties of singly charm pentaquark from decays of $B$ meson,
  including strong coupling constants and production branching ratios of the charm pentaquark. Our numerical results show that the branching ratio of production of pentaquark from B meson can reach to order of $10^{-6}$.
\end{abstract}
\maketitle
  	
\section{Introduction}
Recently, there is resurgent interest in terms of pentaquark, triggered by the progress in the experiment. Since 2015, LHCb collaboration has claimed the discoveries of different pentaquark candidates, hidden charm pentaquark $P_{c}(4380)$, $P_{c}(4312)$, $P_{c}(4440)$, $P_{c}(4457)$, $P_{cs}(4459)$ and $P_{c}(4337)$~\cite{LHCb:2015yax,LHCb:2019kea,LHCb:2020jpq,LHCb:2021chn}. The observations stimulate extensive theoretical studies of pentaquark structures using different scenarios and frameworks~\cite{Santopinto:2016pkp,Deng:2016rus,Maiani:2015vwa,Giron:2021fnl,Lebed:2022vks,Azizi:2022qll,Du:2019pij,Wang:2019ato,Chen:2015loa,Eides:2019tgv,Guo:2015umn,Yang:2022bfu,Wang:2022neq,Qin:2022nof,Richard:2019fms}. In particular, the composite molecular-state interpretation is quite consistent with the LHCb observation. In Ref~\cite{Chen:2019bip, Liu:2019tjn,Shimizu:2019ptd,Wu:2019rog,Lin:2019qiv}, the authors found $P_c(4440)$, $P_c(4457)$ and $P_c(4380)$ can be well interpreted as $\bar D^{(*)}\Sigma_c^{(*)}$ molecules, In Ref~\cite{Peng:2020hql,Xiao:2021rgp,Lu:2021irg,Zhu:2022wpi,Azizi:2023iym}, $P_c(4459)$ and $P_c(4337)$ are believed to be $\bar{D}^* \Xi_c$ and $\Xi_c \bar D$ moleculars respectively. Nevertheless, there are still many other possible proposals to understand the origin of these states, such as compact pentaquark states~\cite{Ali:2019clg,Wang:2019got,Ali:2019npk} and kinematical effects~\cite{Guo:2015umn,Nakamura:2021qvy,Liu:2015fea}. Upon the assumptions, the properties of the pentaquark, such as the various interactions, are investigated with the application of different approaches including the effective approach~\cite{Huang:2018wgr,Lu:2021irg,Zhu:2020jke}, potential models~\cite{Wang:2021hql,Yalikun:2021dpk,Zhu:2021lhd} and QCD sum rules~\cite{Azizi:2022qll,Chen:2015loa,Wang:2022neq,Xin:2023gkf}.
Though there are still many controversial arguments about properties of the hidden charm pentaquark. Moreover, the possible unobserved pentaquark states possessing single charm quark are also quested for with the expectation to be observed in future experiments.

Inspired by the observation of $P_{c}(4337)$ in the analysis of $B_s\to J/\Psi p \bar p$, the present work will focus on the production of pentaquark via B meson decay. Depending on the mass spectrum of singly charm pentaquark based on triquark-diquark model~\cite{Li:2023kcl}, effective Lagrangian method~\cite{Huang:2018wgr,Zhu:2020jke} and Quark Delocalization Color Screening model~\cite{Yan:2023tvl}, we found the masses are around thresholds of pseudoscalar meson baryon, $\Xi D$, $N D_s $ and  $\Sigma D$. In the approaches of QCD sum rules~\cite{Xin:2023gkf} and light-meson exchange model~\cite{Yan:2023ttx}, they are near thresholds of vector meson baryon, $\Xi D^*$, $N D_s^* $ and  $\Sigma D^*$. So it is naturally to follow the assumption that the pentaquark is a meson-baryon molecule state. In order to fully investigate the production of singly charm pentaquark from $B$ meson decay,  we will use the effective Lagrangian approach to describe the interaction of hadrons, and simultaneously discussing both molecule scenarios, including pseudoscalar meson baryon molecular and vector meson baryon molecular in the present consideration.

The building blocks and related evaluation strategy of effective Lagrangian approach to the hadronic molecular, can be sketched out as follows: First, a  gauge invariant phenomenological Lagrangian is established to describes the interaction of the molecular with its constituents. In which, the molecular and constituents are described by standard field operators. By the compositeness condition~\cite{Weinberg:1962hj}, an effective way that representing the probability to find the hadronic molecular as a bare state, the coupling strengths of the hadronic molecular to its constituents can be determined. Using interaction Lagrangian in hadronic level one can further construct transition matrix elements for hadronic processes involving the hadronic molecular. Accordingly, the remaining step of the evaluation of matrix elements will completely reveal the hadronic transition process. We expect our research to provide assistance for the study of singly charm pentaquark in experiments.

The paper is organized as follows. In Sec.~\ref{singly charm pentaquark}, we discuss the possible meson-baryon molecular of pentaquark: pseudoscalar meson baryon $\mathcal{B} P$ and vector meson baryon $\mathcal{B} V$. The coupling constant of pentaquark and its constituent is given in the following. Sec.~\ref{decay amplitude} contains the details of effective Lagrangian approach to calculate the production of singly charmed pentaquark via B meson decay. We present numerical analysis for branching ratios of the production processes in Sec.~\ref{Numerical analysis}. Last section gives a short conclusion.
\section{singly charm pentaquark}
\label{singly charm pentaquark}
The singly charm pentaquark can be decomposed by SU(3) flavor symmetry to form ground 15-plet labeled as $P_c$, in which the states with constituents ${c\bar udss}, {c\bar sudd}, {c\bar udds}, {c\bar duss}, {c\bar suud}$ are exclusively selected as the possible stable pentaquark candidates~\cite{Li:2023kcl}. We arrange $P_c$ to be S-wave states always with spin-parity $J^P={\frac{1}{2}}^{-}$. In previous papers~\cite{Li:2023kcl,Yan:2023ttx,Huang:2018wgr,Zhu:2020jke,Yan:2023tvl,Xin:2023gkf}, the masses of $P_c$ based on different assumptions of their internal structures were studied. It is found that the $J^P={\frac{1}{2}}^{-}$ states with constituents ${c\bar nssn}$, ${c\bar s nnn}$ and ${c\bar nsnn}$($n=u,d$) in the triquark-diquark model~\cite{Li:2023kcl}, effective Lagrangian method~\cite{Huang:2018wgr} and Quark Delocalization Color Screening Model~\cite{Yan:2023tvl} are close to the pseudoscalar meson baryon thresholds $\Xi D$, $N D_s $ and  $\Sigma D$ respectively, indicating existence of a S-wave loosely molecular with meson-baryon $\mathcal{B}P$ components. However, under the QCD sum rules~\cite{Xin:2023gkf} and  light-meson exchange~\cite{Yan:2023ttx}, the masses of ${c\bar s nnn}$ and ${c\bar nsnn}$ with $J^P={\frac{1}{2}}^{-}$ are in proximity to the vector meson baryon thresholds $\Xi D^*$, $N D^*_s $ and  $\Sigma D^*$, suggesting the possibility of $P_{c}$ forming a meson-baryon $\mathcal{B}V$ molecular. So we consider both possible channels relevant to the pentaquark in the work, which explicitly expressed as pseudoscalar meson baryon $\mathcal{B}P$ and vector meson baryon $\mathcal{B}V$.

In general, the mass of the meson-baryon type singly charm pentaquark can be expressed in terms of the constituent meson mass $m_M$, baryon mass $m_{\mathcal{B}}$ and the binding energy $\epsilon$ as,
\begin{eqnarray}
m_{P_{c}}=m_M+m_{\mathcal B}-{\epsilon}.
\end{eqnarray}
Note that $\epsilon$ is a variable quantity in our calculation, here we vary it from 5 to 50 MeV. Once the mass of the composite pentaquark state is fixed, the couplings of pentaquark to meson and baryon can be extracted depending on the compositeness condition. In the following, we discuss the constituent of singly charm pentaquark, respectively regarding as hadronic meson-baryon molecule $\mathcal{B}P$ of $1/2^+$ baryon and pseudoscalar meson interaction or meson-baryon $\mathcal{B}V$ molecular coupled by the interaction $1/2^+$ baryon and vector meson.
\subsection{Meson-baryon S1: $P_c$ as hadronic molecule ${\cal B }P$}
The framework is based on an effective interaction Lagrangian describing the couplings of the pentaquark to their constituents, the nonlocal strong Lagrangian involving $J^P=1/2^-$ pentaquark state($P_c$), a $1/2^+$ baryon($\mathcal{B}$) and pseudoscalar meson($P$) is expressed as follows\,\cite{Lin:2019qiv}:
\begin{eqnarray}
	{\cal L}_{P_{c} {{\cal B}}P}(x) = g_{P_{c} {{\cal B}}P} \, P_{c}(x)
	\int dy \, \Phi(y^2) \, {{\cal B}}^\dagger(x+\frac{y}{2}) \, P(x-\frac{y}{2}) \,.
\end{eqnarray}
The correlation function $\Phi(y^2)$ characterize the finite size of the pentaquark formed as $\mathcal{B}P$ bound states. A basic requirement for the choice of an explicit form of the correlation function is that its Fourier transform vanishes sufficiently fast in the ultraviolet region of Euclidean space to render the Feynman diagrams ultraviolet finite. We adopt the Gaussian form~\cite{Branz:2007xp,Branz:2008cb},
\begin{eqnarray}
\Phi(y^2) \, = \, \int\!\frac{d^4p}{(2\pi)^4}  \,
e^{-ip y} \, {\widetilde{\Phi}}(-p^2) \quad \text{with}\quad
\widetilde\Phi(p^2)
\doteq \exp( - p^2/\Lambda_{P_c}^2),
\end{eqnarray}
where the usual Gaussian regulator cutoff is taken as ${\Lambda_{P_c}}=1.0$ GeV~\cite{Lu:2021irg}. The coupling constant $g_{P_{c} {{\cal B}}P}$ can be easily extracted from the Weinberg compositeness condition~\cite{Weinberg:1962hj}, which implies the renormalization constant $Z_{P_c}$ of the hadron wave function is set equal to zero~\cite{Xiao:2019mvs,Xiao:2016hoa}:
\begin{eqnarray}\label{eq:renormalization}
Z_{P_c} = 1 - g_{P_{c} {{\cal B}}P}^2 \frac{d\Sigma_{P_{c}}(M_{P_{c}}^2)}{dp\!\!\!/}\bigg|_{p\!\!\!/ = m_{P_{c}}} =0 \,.
\end{eqnarray}
The mass operator of pentaquark $\Sigma_{P_{c}}(M_{P_{c}}^2)$ can be described by the self-energy diagram of Fig.\,\ref{fig:selfdiagram}(a), which deduced as,
\begin{flalign}
\Sigma_{P_{c}}(M_{P_{c}}^2)=\int \frac{{\rm d^{4}}k}{(2{\pi})^4}\,\widetilde\Phi^2(-k^2)\frac{({k}\!\!\!/+\frac{p\!\!\!/}{2}+m_{\cal B})}{((k+\frac{p}{2})^2-m^2_{\cal B})((k-\frac{p}{2})^2-m^2_{P})^2}\,,
\end{flalign}
Following the renormalization constant in Eq.(\ref{eq:renormalization}), the coupling constant can be expressed in the form,
\begin{eqnarray}
	\frac{1}{g_{P_{c} P{{\cal B}}}^2} &=& \frac{1}{16\pi^2 }
	\int\limits_0^1 d\alpha
	\, \int\limits_0^\infty \frac{d\beta \beta}{(1 + \beta)^2}
\widetilde\Phi^2(\Delta)\nonumber\\&\times& \Big(\frac{1+2\beta(1-\alpha)}{2(\beta+1)}
+\frac{\beta p\!\!\!/(\beta+1)-{\beta}^2 p\!\!\!/ (2\alpha-1)^2}{{\Lambda_1^2}(\beta+1)}(\frac{p\!\!\!/+2\beta p\!\!\!/(1-\alpha)}{2(\beta+1)}+m_{\cal B})\Big),
\end{eqnarray}
where
\begin{eqnarray}
	\Delta=\beta(m^2_{P}-\alpha m^2_{P}+m^2_{\cal {B}}- \frac{p^2 }{4}+\frac{\beta p^2 }{4+4\beta}-\frac{4\alpha\beta p^2 }{4+4\beta}+\frac{4\alpha^2 \beta p^2 }{4+4\beta}).
\end{eqnarray}
Further determination obtain the numerical result of the couplings, gathered into Tab.\,\ref{tab:bar6}.
\subsection{Meson-baryon S2: $P_c$ as hadronic molecule ${\cal B }V$}
\begin{figure}
  \centering
  \includegraphics[width=0.88\columnwidth]{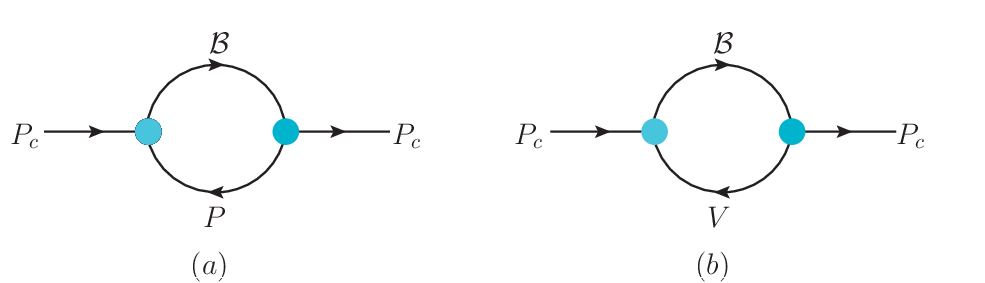}\\
  \caption{The self-energy diagrams of singly charm pentaquark($P_{c}$),  respectively regarded as pseudoscalar meson and baryon molecular $\mathcal B P$(a) or vector meson baryon molecular $\mathcal B V$(b).}\label{fig:selfdiagram}
\end{figure}

The effective interaction Lagrangian for the couplings of $1/2^-$ pentaquark($P_c$), a $1/2^+$ baryon($\mathcal{B}$) and vector meson($V$) is given as \cite{Lu:2021irg}:
\begin{eqnarray}
     {\cal L}_{P_{c}{{\cal B}}V}(x) = g_{P_c{{\cal B}}V} P_{c}(x){\gamma}^{5}(g_{\mu \nu}-\frac{p_{\mu}p_{\nu}}{m^2_{P_{c}}}){\gamma}^{\nu}
	\int dy \, \Phi(y^2) \, {{\cal B}}^\dagger(x+\frac{y}{2}) \, V(x-\frac{y}{2}). \,
\end{eqnarray}
The self-energy diagram of $P_{c}$ as the bound state ${\cal B}V$ is illustrated in Fig.\ref{fig:selfdiagram}(b). Directly the coupling constant can be obtained as:
\begin{eqnarray}
\frac{1}{g_{P_c  \mathcal{B}V}^2}=&&\frac{d}{d p\!\!\!/} \int \frac{d^4
k}{i(2\pi)^4}\ \Phi^2(-k^2)
 \frac{ \gamma^\nu ({k\!\!\!/}+\frac{p\!\!\!/}{2}-m_{\cal B}) \gamma^\rho}{{((k+\frac{p}{2})^2-m^2_{\cal B})((k-\frac{p}{2})^2-m^2_{V})}}\nonumber\\
&&(g_{\mu\nu}-\frac{p_{\mu} p_{\nu}}{m_{P_{c}}^2})(g_{\sigma\rho}-\frac{p_{\sigma} p_{\rho}}{m_{P_{c}}^2})({-g^{\mu\sigma}+\frac{(k-\frac{p}{2})^{\mu}(k-\frac{p}{2})^{\sigma}}{m_{V}^2}})
\end{eqnarray}
where $k+\frac{p}{2}$ is the momentum of baryon ${\cal B}$ and $k-\frac{p}{2}$ is the momentum of vector meson $V$ respectively.
\section{The production from B meson decay}\label{decay amplitude}
\begin{figure}
  \centering
  \includegraphics[width=0.77\columnwidth]{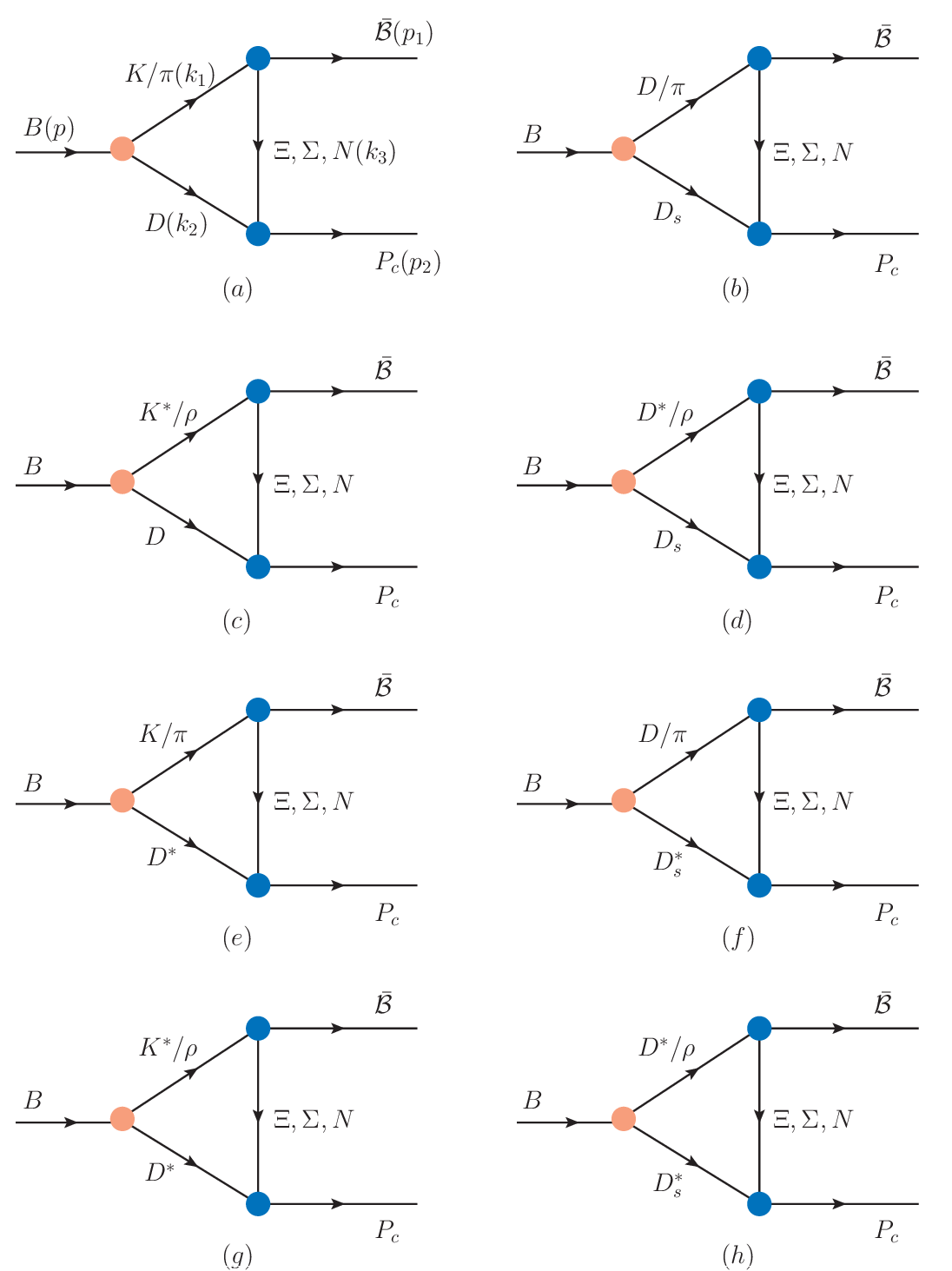}
  \caption{The triangle diagrams of singly charm pentaquark produced by two body decays of B meson. In (a-d), pentaquark is assumed to be pseudoscalar meson baryon $\mathcal B P$ molecular, while in (e-h) it exists as vector meson baryon $\mathcal{B} V$ molecular. The exchanged particles can be light baryons $N$, $\Xi$ and $\Sigma$.}\
\label{fig:triangle}
\end{figure}

\begin{figure}
  \centering
  \includegraphics[width=0.55\columnwidth]{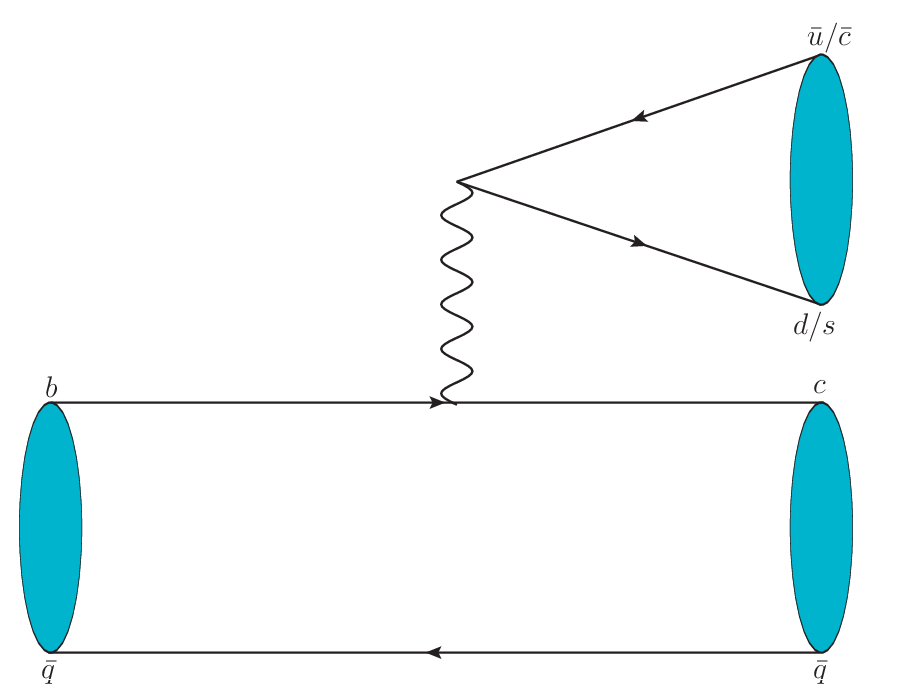}\\
  \caption{The dominant contribution of weak decay of B meson coming from the factorizable W-emission diagram.}\label{fig:Wemission}
\end{figure}

Using the effective Lagrangian method, we could further study the production properties of pentaquark from B meson decays. The naive two body production $B\to P_{c} {\bar{\mathcal B}}$ can be divided into two subprocesses under the phenomenological approach, by inserting possible hadron complete states, expressed as a weak transition matrices $\langle\lambda|\mathcal{H}_{eff}| B\rangle$ and a strong coupling matrices $\langle P_{c} \overline {\mathcal{B}}|\mathcal{H}_{\lambda}|\lambda\rangle$.
\begin{eqnarray}
\langle P_{c} \overline {\mathcal{B}}|\mathcal{H}_{eff}| B\rangle=\sum_{\lambda}\langle P_{c} \overline {\mathcal{B}}|\mathcal{H}_{\lambda}|\lambda\rangle\langle\lambda|\mathcal{H}_{eff}| B\rangle,
\end{eqnarray}
where $\overline {\mathcal{B}}$ can be charm anti-baryon ${\bar c \bar q \bar q}$ or light anti-baryon ${\bar q \bar q \bar q}$. The possible intermediate hadron states($\lambda$) of  charm-light mesons and charm-charm mesons ($\lambda=D^{(*)} M,\ D^{(*)}D$) have the dominant contribution, in light of the leading factorizable process of weak decays of B meson shown in Fig.\ref{fig:Wemission}. $H_{\lambda}$ is the possible Lagrangian of strong interaction in hadronic level, which wound be discussed in the following. Under the framework of SU(3) light quark symmetry~\cite{Shi:2017dto,Xing:2018bqt,Xing:2022aij}, we could select several golden production channels, corresponding to quark  transitions $b\to c \bar c d/s$ and $b\to  c \bar u d/s$~\cite{Li:2023kcl},
\begin{eqnarray}
\overline B_s^0\to   P_{c\bar suud}^{++}  \overline \Sigma^{--}_{\bar c},\,
\overline B^0\to   P_{c\bar udss}^-  \overline \Xi^{+},\,
\overline B_s^0\to   P_{c\bar sudd}^+  \overline p,\,
B^-\to   P_{c\bar udds}^-  \overline \Lambda^{0},\,
\overline B^0\to   P_{c\bar duss}^+  \overline \Sigma^{-}.
\end{eqnarray}
The processes can be described by chain $B\to M_1 M_2 \xrightarrow[]{{\mathcal{B}}} P_{c}\overline{\mathcal{B}}$ on account of effective Lagrangian method, pictured as triangle diagrams in Fig.~\ref{fig:triangle}. Here, two intermediate mesons $M_1 M_2$ can be charm-light mesons $D^{(*)} M$ or charm-charm mesons $D^{(*)}D$, they exchange by baryon($\mathcal{B}$) to produce the expected pentaquark.
\subsection{Effective Lagrangian and Form Factors}
\label{Effective Lagrangian}
To describe the chain process $B\to M_1M_2$, the effective Hamiltonian of weak decays of B meson is ready
\begin{eqnarray}
\mathcal{H}_{eff} &=& \frac{G_F}{\sqrt{2}} V_{CKM} (C_1(\mu) \mathcal{O}_1(\mu)+C_2(\mu) \mathcal{O}_2(\mu))+h.c. \,
\end{eqnarray}
Here $G_F$ is the Fermi constant, $C_{1,2}(\mu)$ and $O_{1,2}(\mu)$ are the tree level Wilson coefficients and fermion operators of transition $b\to c\bar u d/s$ or $b\to c\bar c d/s$, for instance, $\mathcal{O}_{1}=(\bar c_{\alpha} b_{\beta})_{V-A}(\bar d_{\beta}u_{\alpha})_{V-A}$, $\mathcal{O}_{2}=(\bar c_{\alpha} b_{\alpha})_{V-A}(\bar d_{\beta}u_{\beta})_{V-A}$ for the transition $b\to c \bar u d$. The weak transition matrices $\langle\lambda|\mathcal{H}_{eff}| B\rangle$ can be further expressed as
\begin{eqnarray}
&&\mathcal{M}( B \to D^{({*})} M)=\frac{G_F}{\sqrt{2}} V_{CKM}\, a_1  \langle D^{({*})}|(\bar c b)_{V-A} | B\rangle \langle M|(\bar qq)_{V-A} |0\rangle,
\end{eqnarray}
where ${\bar q q}=\bar d u$, $\bar s u$ or $\bar s c$. $a_1$ appears as the combinations of effective Wilson coefficient $a_1=C_1+C_2/N_c$ with $N_c$ the number of colors.

The parameterized forms of $B \to D$ and $B \to D^*$ are~\cite{Penalva:2023snz,McLean:2019qcx,Harrison:2021tol}
\begin{align}
\langle D(k_{2}) | (\bar cb)_{V-A} |B(p)\rangle&= F_{1}({k^2_1})(p^{\mu}+k_{2}^{\mu}-\frac{m_{B}^2-m_{D}^2}{k^2_1}k_1^{\mu}) +F_{2}(k^2_1)(\frac{m_{B}^2-m_{D}^2}{k^2_1}k^{\mu}_1),\nonumber\\
\langle D^*(k_{2}) | (\bar cb)_{V-A} |B(p)\rangle&=\frac{2iA_0(k^2_1)}{m_{B}+m_{D^*}}{\varepsilon}^{{\mu}{\nu}{\rho}{\sigma}}{\epsilon}^*_{\nu}{k_{2}}_{\rho}{p}_{\sigma}\nonumber
                                  -2m_{D^*}A_{1}(k^2_1)\frac{{{\varepsilon}^*}(k_{2})\cdot k_1}{k^{2}_1}k^{\mu}_1\nonumber\\
                                  &-(m_{B}+m_{D^*})A_{2}(k^2_1)({{\varepsilon}^*}^{\mu}(k_{2})-\frac{{{\varepsilon}^*}(k_{2})\cdot k_1}{k^{2}_1}k^{\mu}_1)\nonumber\\
                                  &+A_{3}(k^2_1)\frac{{{\varepsilon}^*}(k_{2})\cdot k_1}{m_{B}+m_{D^*}}(p^{\mu}+k^{\mu}_{2}-\frac{m^2_{B}-m^2_{D^*}}{k^2_1}k^{\mu}_1).
\end{align}
Where $p$ and $k_2$ represent momentum of $B$ and $D^{(*)}$ mesons. $\varepsilon^{*}$ denotes the polarization vector of vector meson $D^*$, and $k_1=p-k_2$ is transition momentum.

The matrix elements involving pseudoscalar or vector mesons and the vacuum exhibit the following definitions~\cite{Cheng:1996cs},
\begin{eqnarray}
\langle V(k_1,\varepsilon_{\mu}^*)| (\bar qq)_{V-A}|0 \rangle=m_V f_{V} \varepsilon^*_{\mu}, \qquad \langle P(k_1)|(\bar qq)_{V-A} |0\rangle=i f_{P} \, {k_1}_{\mu}.
\end{eqnarray}
Wherein $f_{P}$ and $f_{V}$ are decay constants of pseudoscalar $P$ and vector $V$ mesons  respectively.

The description about chain $M_1M_2\to P_c \bar{\mathcal{B}}$ can be accomplished by the effective Lagrangian including two baryons and a meson, given as~\cite{Yalikun:2021dpk}:
\begin{eqnarray}
\mathcal{L}_{\mathcal{B}\mathcal{B}P} &=& \frac{g_{\mathcal{B}\mathcal{B}P}}{m_{P}}{} \langle\overline{\mathcal{B}} \gamma_5\gamma_{\mu} \partial^{\mu}P \mathcal{B}\rangle,\\
\mathcal{L}_{\mathcal{B}\mathcal{B}V} &=& g_{\mathcal{B}\mathcal{B}V}{} \langle\overline{\mathcal{B}} \gamma_{\mu} V^{\mu} \mathcal{B}\rangle+ \frac{f_{\mathcal{B}\mathcal{B}V}}{2 m_V} \langle \overline{\mathcal{B}} \sigma_{\mu\nu} \partial^{\mu}V^{\nu} \mathcal{B} \rangle.
\end{eqnarray}
The coupling constants between meson and two baryons can be determined by the generic SU(4) symmetry, we deduce them into Tab.\ref{tab:bar61}.
\subsection{Decay width}
Armed with the effective Lagrangian and form factor, the production process of pentaquark is direct. We show amplitudes of process $\overline B^0_s\to P^+_{c\bar s udd} \bar p$ corresponding with Fig.\ref{fig:triangle}(b,d,f,h), given as
\begin{eqnarray}
{\mathcal M}_b \!\!&=&i f_{\pi}\!\! \int \frac{d^4k_{\text{\tiny $3$}}}{(2\pi)^4}g_{\text{\tiny $NN \pi$}}g_{\text{\tiny $P_{c}N D_s$}} \frac{\bar u_{\text{\tiny $P_{c}$}}(p_{\text{\tiny $2$}})(/\!\!\!k_{\text{\tiny $3$}}+m_{\text{\tiny $N$}})\gamma^{5}{k\!\!\!/}_{1}\nu_{\text{\tiny $\overline p$}}(p_{\text{\tiny $1$}})}{(k_{\text{\tiny $1$}}^2-m_{\text{\tiny $\pi$}}^2)\ (k_{\text{\tiny $2$}}^2-m_{\text{\tiny $D_s$}}^2)\ (k_{\text{\tiny $3$}}^2-m_{\text{\tiny $N$}}^2)}\mathcal{F}^2(k_{\text{\tiny $3$}}^2)\nonumber\\&&
 \Big((2p_{\text{\tiny $2$}}+p_{\text{\tiny $1$}}-k_{\text{\tiny $3$}})\cdot (p_{\text{\tiny $1$}}+k_{\text{\tiny $3$}}) F_1(k_{\text{\tiny $1$}}^2)-(m^2_{\text{\tiny $B$}}-m^2_{\text{\tiny $D_s$}})F_1(k_{\text{\tiny $1$}}^2)+(m^2_{\text{\tiny $B$}}-m^2_{\text{\tiny $D_s$}})F_2(k_{\text{\tiny $1$}}^2)\Big),\\
{\mathcal M}_d \!\!&=&f_{\rho}m_{\rho}\!\! \int \frac{d^4k_{\text{\tiny $3$}}}{(2\pi)^4}g_{\text{\tiny $NN \rho$}}g_{\text{\tiny $P_{c}ND_s$}} \frac{\bar u_{\text{\tiny ${P_c}$}}(p_{\text{\tiny $2$}})(/\!\!\!k_{\text{\tiny $3$}}+m_{\text{\tiny $N$}})(\gamma_{\nu}-f_{NN \rho}\sigma_{\xi \nu }{k_{\text{\tiny $1$}}^{\xi}}/4m_{\rho})\nu_{\text{\tiny $\overline p$}}(p_{\text{\tiny $1$}})}{-(k_{\text{\tiny $1$}}^2-m_{\text{\tiny $\rho$}}^2)\ (k_{\text{\tiny $2$}}^2-m_{\text{\tiny $D_s$}}^2)\ (k_{\text{\tiny $3$}}^2-m_{N}^2)}\nonumber\\
&&(-g^{\nu \mu}+\frac{{k^{\nu}_1} {k^{\mu}_1}}{m_{\rho}^2}) \mathcal{F}^2(k_{\text{\tiny $3$}}^2)
 \Big({(p+{k_2})}_{\mu}F_1(k_{\text{\tiny $1$}}^2)\Big),\\
{\mathcal M}_{f}\!\!&=&if_{\pi}\!\!\int\frac{d^{4}k_{3}}{( 2\pi) ^{4}}g_{\text{\tiny $NN \pi$}}g_{\text{\tiny $P_{c}N{D^*_s}$}}\frac{\bar u_{\text{\tiny $P_{c}$}}( p_{2})\gamma^{\beta}(/\!\!\!k_3-m_{N} ){k\!\!\!/}_{1} {\nu}_{\bar p}(p_{1} ) (g_{\alpha\beta}-\frac{{p_{2}}_{\alpha}{p_{2} }_{\beta}} {m^{2}_{P_{c}}}) (-g^{\alpha\mu}+\frac{k^{\mu}_{2}k^{\alpha}_{2}}{m^{2}_{D^*_{s}}}) }{(k^{2}_{1}-m^{2}_{\pi} )(k^{2}_{2}-m^{2}_{D^{*}_{s}})(k^{2}_{3}-m^{2}_{N} )  }\nonumber\\&&
\Big(-2m_{D^*_{s}}A_{1}(k^{2}_{1}){k_{1}}_{\mu}
+A_{3}( k^{2}_{1}) \frac{{k_{1}}_{\mu}}{m_{B}+m_{D^*_{s}}}(p\cdot k_{1}+k_{2}\cdot k_{1}-m^{2}_{B}+m^{2}_{D^*_{s}})\Big)\mathcal{F}^2(k_{\text{\tiny $3$}}^2),\\
{\mathcal M}_{h}\!\!&=&f_{\rho}m_{\rho}\!\!\int\frac{d^{4}k_{3}}{2\pi ^{4}}g_{\text{\tiny $NN \rho$}}g_{\text{\tiny $P_{c}N{D^*_s}$}}\frac{\bar u_{\text{\tiny $P_{c}$}}( p_{2}) \gamma_{5}\gamma^{\beta}(/\!\!\!k_3+m_{N} )(\gamma_{\delta}-f_{NN \rho}\sigma_{\eta \delta}k_{\text{\tiny $1$}}^{\eta}/4m_{\rho}) \nu_{\text{\tiny $\overline p$}}(p_{\text{\tiny $1$}}) }{-(k^{2}_{3}-m^{2}_{N} )(k^{2}_{2}-m^{2}_{D^{*}_{s}} )(k^{2}_{1}-m^{2}_{\rho})}\nonumber\\&&
\Big(\frac{2iA_0(k^{2}_{1} ) }{m_{B}+m_{D^*_{s}}}\varepsilon_{\mu\rho\sigma \nu}{ k^{\rho}_{2}}p ^{\sigma}-A_{2}( k^{2}_{1}) g_{\nu\mu}
(m_{B}+m_{D^*_{s}})+A_{3}( k^{2}_{1})\frac{{ k_{1}}_{\nu} }{m_{B}+m_{D^*_{s}}}(p+k_{2} )_{\mu}\Big)\nonumber\\
&&\mathcal{F}^2(k_{\text{\tiny $3$}}^2)(g_{\alpha\beta}-\frac{{p_2}_{\alpha}{p_2}_{\beta}}{m^{2}_{P_{c}}} )(-g^{\alpha\nu}+\frac{{k^{\alpha}_2}{k^{\nu}_2}}{m_{D^*_{s}}})(-g^{\mu\delta}+\frac{{k^{\mu}_1}{k^{\delta}_1}}{m^{2}_{\rho}}).
\end{eqnarray}
Where $m_{N}$ and $k_{3}$ represent the mass and momentum of exchanged baryon $N$. The momentum definitions can be found in Fig.\ref{fig:triangle}. To describe the structure and off-shell effect, a monopole form factor is introduced for the exchanged baryon ${\cal B}$\,($\Xi,\,\Sigma,\,N$)~\cite{Cheng:2004ru,Tornqvist:1993ng},
\begin{eqnarray*}
\mathcal{F}(k_3^2)=\frac{{m_{E}^2}-\Lambda^2}{k_3^2-\Lambda^2}.
\end{eqnarray*}
$m_{E}$ and $k_{3}$ represent the mass and momentum of exchanged baryons ${\cal B}$. We adopt the cutoff value $\Lambda=m_{E}+\alpha$\,\cite{Xing:2022aij,Cheng:2004ru,Tornqvist:1993vu,Tornqvist:1993ng,Locher:1993cc,Li:1996yn}.

The two-body decay width of $B$ meson ${B}\to P_{c} \bar{\cal {B}}$ can be formulated as
\begin{eqnarray*}
\Gamma(B\to P_{c}+\overline {\mathcal{B}})=\frac{|\mathbf{P_f}|}{8\pi m_{B}^2}|\mathcal{M}|^2,
\end{eqnarray*}
here, $\mathcal{M}$ represents amplitude of singly charm pentaquark produced from $B$ meson, and $\mathbf{P_f}$ is the magnitude of the three-momentum of final state, known as $|\mathbf{P_f}|=\frac{1}{2m_{B}}\sqrt{\lambda(m_{B}^2,m^2_{P_{c}},m^2_{\overline {\mathcal{B}}})}$, with $\lambda(a,b,c)=a^2+b^2+c^2-2ab-2bc-2ac$.
\section{Numerical analysis}
\label{Numerical analysis}
\begin{figure}
\includegraphics[width=0.95\columnwidth]{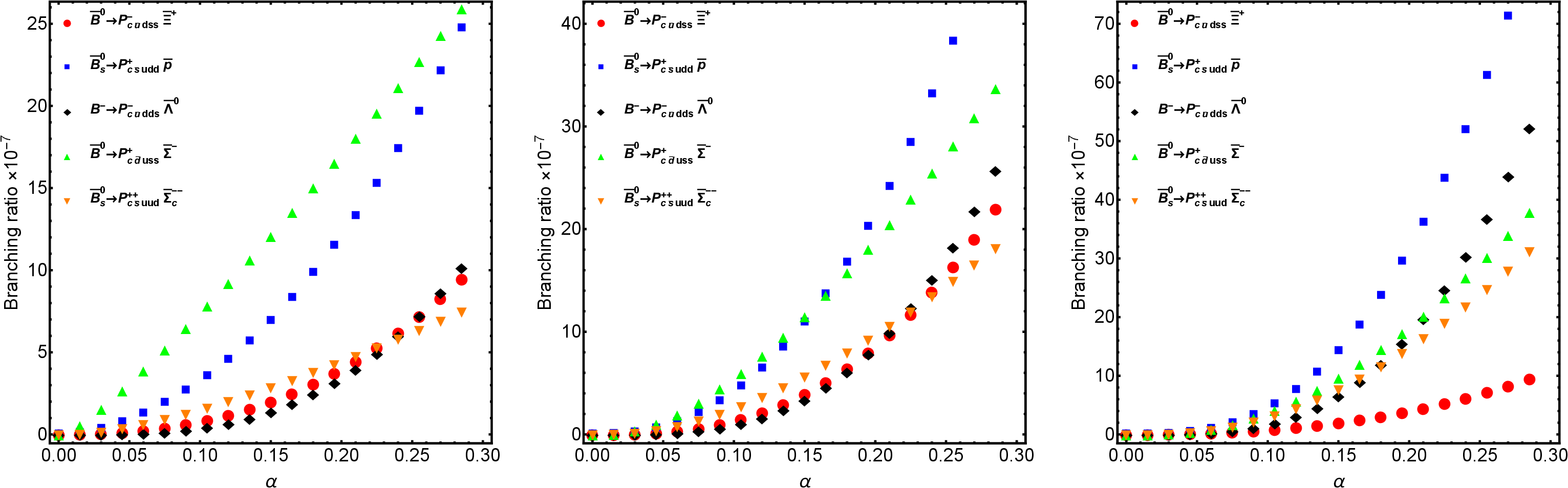}
\caption{In the configuration of meson-baryon $\mathcal{B}P$ molecular(S1), the branching ratios of ${B}\to {P}_{c} \overline {\mathcal{B}}$ vary with the cutoff parameter $\alpha$. The graphs correspond with binding energy $\epsilon$=5 MeV(left), 20 MeV(middle) and 50 MeV(right).}
\label{fig:BrBP}
\end{figure}

\begin{figure}
\includegraphics[width=0.95\columnwidth]{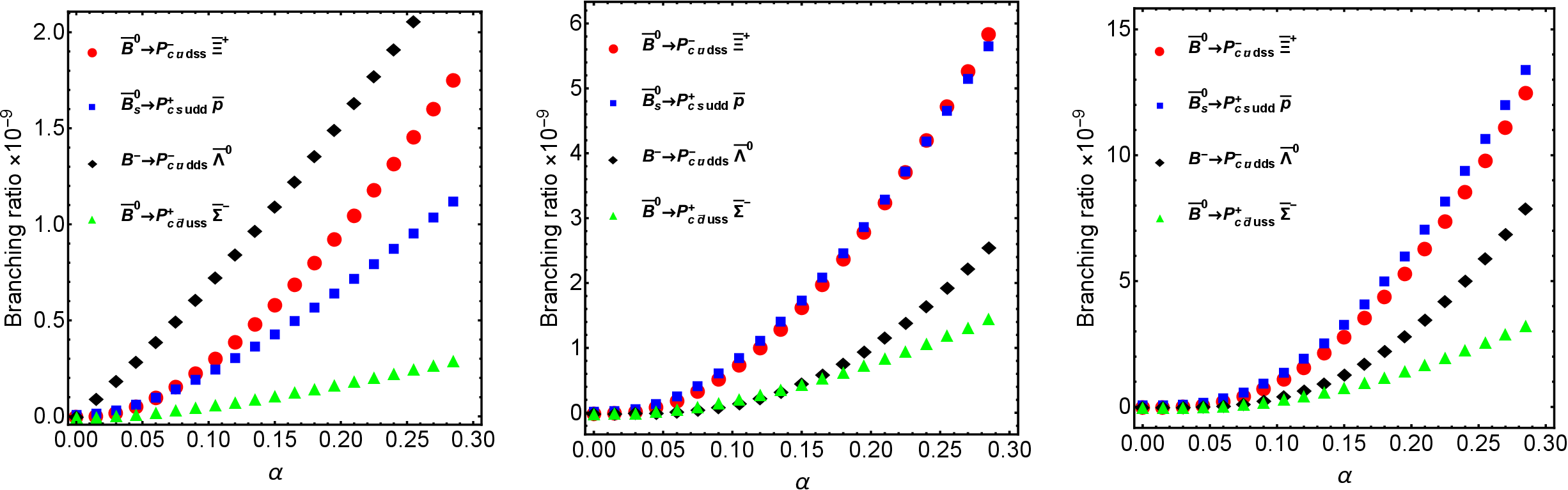}
\caption{In the configuration of meson-baryon $\mathcal{B}V$ molecular(S2), the branching ratios of ${B}\to {P}_{c} \overline {\mathcal{B}}$ vary with the cutoff parameter $\alpha$. The graphs correspond with binding energy $\epsilon$=5 MeV(left), 20 MeV(middle) and 50 MeV(right).}
\label{fig:BrBV}
\end{figure}

In the numerical analysis, the form factors $F_{1}(k^2)$, $F_{2}(k^2)$ and $A_i(k^2)$($i$=1,2,3) of transition $B\to D(D^*)$ can be parameterized using the Bourrely-Caprini-Lellouch (BCL) method~\cite{ McLean:2019qcx,Harrison:2021tol}
\begin{eqnarray}
&F_{1}(k^2)&=\frac{1}{1-k^2/m^2_{B_{c}}}(a_{0}z^0+a_{1}z^1+a_{2}z^2),\nonumber\\
&F_{2}(k^2)&=\frac{1}{1-k^2/m^2_{B^*_c}}\big(a_{0}z^0+a_{1}(z^1-\frac{z^3}{3})+a_{2}(z^2+\frac{2z^3}{3})\big),\nonumber\\
&A_i(k^2)&=(\frac{\sqrt{(m_{B}+m_{D^*})^2-k^2}+\sqrt{(m_{B}+m_{D^*})^2-m^2_{{pole}}}}{\sqrt{(m_B+m_{D^*})^2-k^2}-\sqrt{(m_B+m_{D^*})^2-m^2_{{pole}}}})(a_{0}+a_{1}z_1+a_{2}z_1^2+a_{3}z_1^3),
\end{eqnarray}
with the expressions of $z$- and $z_1$-coefficient
\begin{eqnarray}
&z=\frac{\sqrt{(m_{B}+m_{D})^2-k^2}-\sqrt{(m_{B}+m_{D})^2}}{\sqrt{(m_{B}+m_{D})^2-k^2}+\sqrt{(m_{B}+m_{D})^2}},\
z_1=\frac{\sqrt{(m_B+m_{D^*})^2-k^2}-\sqrt{(m_B+m_{D^*})^2-(m_{B_s}-m_{D^*_s})^2}}{\sqrt{(m_B+m_{D^*})^2-k^2}+\sqrt{(m_B+m_{D^*})^2-(m_{B_s}-m_{D^*_s})^2}}.
\end{eqnarray}
Where $m_{pole}$ denotes the physical pole masse of $B_c$ meson. The fitted parameters $a_{0,1,2,3}$ and $m_{pole}$ are collected into Tab.~\ref{tab:bar6}. Additionally, we need decay constants of mesons~\cite{Yalikun:2021dpk},
\begin{eqnarray*}
  &f_{\pi}=0.130\, \text{GeV},\,\,f_K=0.156\, \text{GeV},\,\,f_D=0.212\, \text{GeV},\,\,f_{D_s}=0.249\, \text{GeV},\,\,\\
&f_{K^*}=0.210\, \text{GeV},\,\,f_{D^*}=0.220\, \text{GeV},\,\,f_{\rho}=0.216\, \text{GeV},\,\,f_{D_s}=0.230\, \text{GeV},
\end{eqnarray*}
as well as CKM matrix elements and effective Wilson coefficient~\cite{Li:2012cfa,Xing:2019xti},
\begin{eqnarray*}
&&|V_{cd}|=0.221,\,\,|V_{ud}|=0.974,\,\,|V_{us}|=0.2243,\,\,|V_{cd}|=0.221,\,\,|V_{ud}|=0.974,\\
&&G_F=1.166 \times10^{-5},\,\,
a_{1}=1.07,\,\,{\tau}_{B}=1.638\times10^{-12}s.
\end{eqnarray*}
\begin{table}
\caption{The fitted expansion parameters $a_i$ of form factors for transitions ${B\to D}$ and ${B\to D^*}$.}\label{tab:bar6}
\begin{tabular}{c |c c c c c c c c c}\hline\hline
&&  & $a_0$&$a_1$ &$a_2$&$a_3$&$m_{pole}$(GeV) \\\hline
\multirow{2}{*}{${B\to D}$}&
&$F_1(0)$& $0.666$& $-0.260$& $-0.106$& $0.00$&$-$ \\
&&$F_{2}(0)$& $0.666$ &
$-3.236$& $-0.075$ &
$0.00$&$-$
\\\hline
\multirow{4}{*}{${B\to D^*}$}&
&$A_{0}(0)$& $0.100$ &
$-0.180$& $-0.006$ &
$0.00$&$6.335$ \\
&&$A_{1}(0)$& $0.105$ &
$-0.430$& $-0.100$ &
$-0.03$&$6.275$ \\
&&$A_{2}(0)$& $0.055$ &
$-0.010$& $-0.030$ &
$0.06$&$6.745$ \\
&&$A_{3}(0)$& $0.059$ &
$-0.110$& $-0.250$ &
$-0.05$&$6.745$
\\\hline
\end{tabular}
\end{table}
\begin{table}
\centering
\caption{The coupling constants of singly charm pentaquark and the molecular meson-baryon components, obtained by compositeness condition. The couplings between meson and two baryons, derived from SU(4) flavor symmetry, where the generic coupling parameters are taken as $g_{\mathcal{B}\mathcal{B}P}=0.989$, $g_{\mathcal{B}\mathcal{B}V}=3.25$ and $f_{\mathcal{B}\mathcal{B}V}=6.1$~\cite{Yalikun:2021dpk}.} \label{tab:bar61}
\begin{tabular}{c c c c c c c c c c c c}\hline\hline
$g_{\text{\tiny{$P_{c }\Xi D$}} }$&$g_{\text{\tiny{$P_{c}ND_s$}} }$&$g_{\text{\tiny{$P_{c}\Sigma D$}}}$&$g_{\text{\tiny{$P_{c }\Xi D^*$}} }$&$g_{\text{\tiny{$P_{c}ND^*_s$}} }$&$g_{\text{\tiny{$P_{c}\Sigma D^*$}}}$ \\
2.280&2.209&2.282&0.337&0.293&0.328&\\
$g_{\text{\tiny{${\Xi\Xi \pi}$}}}$&$g_{\text{\tiny{${\Sigma \Lambda \pi}$}}}$&$g_{\text{\tiny{${\Sigma_c N D}$}}}$&$g_{\text{\tiny{${\Xi\Sigma K}$}}}$&$g_{\text{\tiny{${NN\pi}$}}}$&$g_{\text{\tiny{${\Xi \Xi_c D_s}$}}}$\\
3.78&9.97&5.34&19.09&13.5&13.41\\
$g_{\text{\tiny{${NN\rho}$}}}$&$g{\text{\tiny{$_{\Xi \Xi_c D^*_s}$}}}$& $g_{\text{\tiny{${\Xi\Sigma K^*}$}}}$& $g_{\text{\tiny{${\Xi\Xi \rho}$}}}$& $g_{\text{\tiny{${\Lambda\Sigma \rho}$}}}$&$g{\text{\tiny{$_{\Sigma_c ND^*}$}}}$\\
3.25&1.88&4.60&3.25&3.75&4.60\\
$f_{\text{\tiny{${NN\rho}$}}}$&$f{\text{\tiny{$_{\Xi \Xi_c D^*_s}$}}}$& $f_{\text{\tiny{${\Xi\Sigma K^*}$}}}$& $f_{\text{\tiny{${\Xi\Xi \rho}$}}}$& $f_{\text{\tiny{${\Lambda\Sigma \rho}$}}}$&$f{\text{\tiny{$_{\Sigma_c ND^*}$}}}$\\
6.1&3.52&8.63&6.1&7.04&8.63
\\\hline
\end{tabular}
\end{table}
We adopt cutoff parameter $\alpha$ with a central value 250 MeV to determine the branching ratios of charm pentaquark $P_{c}$ produced from $B$ meson, respectively at the binding energy ${\epsilon}=5, 20$ and $50$ MeV. The branching ratios are achieved at the order of $10^{-7}\sim10^{-6}$ for the meson-baryon scenario $\mathcal{B}P$ molecular, and $10^{-10}\sim10^{-9}$ for the meson-baryon $\mathcal{B}V$ molecular, the results are shown in Tab.~\ref{tab:ratio}.
For completeness, we draw branching ratios changing with cutoff parameter $\alpha$ in range of $100-300$ MeV in Fig.~\ref{fig:BrBP} and Fig.~\ref{fig:BrBV}.

Among the processes with attention, the branching ratio under pentaquark as hadronic molecule $\mathcal{B}P$ (S1) holds the higher possibility than the meson-baryon $\mathcal{B}V$ molecular (S2). It is noteworthy that the process $\overline B_s^0\to   P_{c \bar sudd}^{++}  \overline \Sigma^{--}_{\bar c}$ is unallowed under the latter molecular case owing to the effect of phase space. The results of branching ratios obviously show that, the production channels $\overline B^0_s\to   P_{c \bar sudd}^{+}  \overline p$ and $\overline B^0 \to P_{c\bar duss}^+ \overline{\Sigma}^{-}$ possess considerable order of branching ratio in two molecular scenarios, for instance, in $\mathcal{B}P$ molecular at the binding energy $\epsilon$=20MeV,
\begin{eqnarray}
&&\text{Br}(\overline B^0_s\to   P_{c \bar sudd}^+  \overline p)=3.648\times10^{-6},\  \text{br}(\overline B^0_s\to \rho D_s \xrightarrow[]{N}P^+_{c\bar s udd} { \overline p})=2.59\times 10^{-6},\nonumber\\
&&\text{Br}(\overline B^0 \to P_{c\bar duss}^+ \overline{\Sigma}^{-})=2.727\times10^{-6},\ \text{br}(\overline B^0\to K D \xrightarrow[]{\Xi}P^+_{c\bar duss} { \overline \Sigma}^-)=2.60\times 10^{-6}.
\end{eqnarray}
In particularly, the dominant contributions among the two possible sub processes come from $\overline B^0_s\to \rho D_s \xrightarrow[]{N}P^+_{c\bar s udd} { \overline p}$ as Fig.~\ref{fig:triangle}.(d) and $\overline B^0\to K D \xrightarrow[]{\Xi}P^+_{c\bar duss} { \overline \Sigma}^-$ as Fig.~\ref{fig:triangle}.(a).
\begin{table}
  \centering
  \caption{The production processes of pentaquark(${B}\to {P}_{c} \overline {\mathcal{B}} \, $) in B meson decays can be discussed separately in two scenarios, S1: pentaquark as the hadron molecule of $\mathcal{B}P$, S2: pentaquark as the hadron molecule of $\mathcal{B}V$. The branching ratios are calculated at a central value $\Lambda=m_{E}+0.25$ GeV, within the value of binding energy ${\epsilon}=5$, $20$, $50$ MeV. The exchanged particles (EP) can be $\Xi,N, \Sigma$. For definiteness, we show the dominant intermediate hadron states in the last column.}\label{tab:ratio}
       \begin{tabular}{|c|c|c|c|c|c|c|c}\hline
      \multirow{2}*{}& \multirow{2}*{${B}\to {P}_{c} \overline {\mathcal{B}}$} &  \multicolumn{3}{|c|}{Branching ratio($\times 10^{-7}$)} 	& \multirow{2}*{EP} & \multirow{2}*{Dominant hadrons}  \\ \cline{3-5}
    &  &  $\epsilon=5$MeV\,\,\,\,\,\,\,  & $\epsilon=20$MeV\,\,\,\,\,&$\epsilon=50$MeV & &    \\\hline
 &$\overline B^0\to   P_{c \bar udss}^-  \overline \Xi^{+}$  &6.84 &$15.49$ &28.51& $\Xi$& $\rho D$\\
 &$\overline B_s^0\to   P_{c \bar sudd}^+  \overline p$  &18.88 &$36.48$ &57.92& $N$& $\rho D_s$\\
S1 &$ B^-\to   P_{c \bar udds}^-  \overline \Lambda^{0}$  &6.80 &$17.16$ &34.52& $\Sigma$& $\rho D$\\
&$\overline B^0\to   P_{c \bar duss}^+  \overline \Sigma^{-}$ &22.20 &$27.27$ &29.06& $\Xi$& $K D$\\
&$\overline B_s^0\to   P_{c \bar suud}^{++}  \overline \Sigma^{--}_{\bar c}$ &6.17 &$14.43$ &23.74&  $N$&$DD_s$\\\hline
 &$\overline B^0\to   P_{c \bar udss}^-  \overline \Xi^{+}$  &0.014 &$0.045$ &0.094& $\Xi$& $\rho D^*$\\
 &$\overline B_s^0\to   P_{c \bar sudd}^+  \overline p$  &0.009 &$0.045$ &0.100& $N$& $\rho D^*_s$\\
S2 &$ B^-\to   P_{c \bar udds}^-  \overline \Lambda^{0}$  &0.020 &$0.018$ &0.056& $\Sigma$& $\rho D^*$\\
&$\overline B^0\to   P_{c \bar duss}^+  \overline \Sigma^{-}$  &0.002 &$0.012$ &0.025& $\Xi$& $K D$,\,$K^* D$
\\\hline
    \end{tabular}
\end{table}
\section{Conclusions}
In this work, we force on the properties and productions of singly charm pentaquark. Under the assumption of pentaquark molecular composition, it is meaningful to investigate the productions of singly charm pentaquark by decays of B meson ${B}\to {P}_{c} \overline {\mathcal{B}}$. According to the effective Lagrangian method, the selected suitable processes can then be studied with the benefit of hadronic triangle diagrams. We deduce branching ratios of the productions eventually, what deserves our attention is the productions of pentaquark $P_{c\bar sudd}^+$ and $P_{c\bar d uss}^+$. Our calculations suggest that branching ratios of the two Cabibbo allowed processes can reach to $10^{-6}$ and $10^{-8}$, which needs to be further confirmed by experiments.
\section*{Acknowledgments}
Y.X. acknowledges the support by National Natural Science Foundation of China under Grant No. 12005294, and the Fundamental Funds for Key disciplines in Physics with No.2022WLXK05.
\appendix
\section{Decay amplitudes}\label{sec:appenda}
The amplitude of $\overline B^0\to P_{c\bar udss}^-  \overline \Xi^{+}$ is
\begin{eqnarray}
{\cal M}_a \!\!&=&i f_{\pi}\!\! \int \frac{d^4k_{\text{\tiny $3$}}}{(2\pi)^4}g_{\text{\tiny $\Xi \Xi \pi$}}g_{\text{\tiny $P_{c}\Xi D $}} \frac{\bar u_{\text{\tiny $P_{c}$}}(p_{\text{\tiny $2$}})(/\!\!\!k_{\text{\tiny $3$}}+m_{\text{\tiny $\Xi$}})\gamma^{5}{k\!\!\!/}_{1}\nu_{\text{\tiny $\overline \Xi$}}(p_{\text{\tiny $1$}})}{(k_{\text{\tiny $1$}}^2-m_{\text{\tiny $\pi$}}^2)\ (k_{\text{\tiny $2$}}^2-m_{\text{\tiny $D$}}^2)\ (k_{\text{\tiny $3$}}^2-m_{\text{\tiny $\Xi$}}^2)}\nonumber\\&&\mathcal{F}^2(k_{\text{\tiny $3$}}^2)
 \Big((2p_{\text{\tiny $2$}}+p_{\text{\tiny $1$}}-k_{\text{\tiny $3$}})\cdot (p_{\text{\tiny $1$}}+k_{\text{\tiny $3$}})-m^2_{\text{\tiny $B$}}+m^2_{\text{\tiny $D$}})F_1(k_{\text{\tiny $1$}}^2)+(m^2_{\text{\tiny $B$}}-m^2_{\text{\tiny $D$}})F_2(k_{\text{\tiny $1$}}^2)\Big),\\
{\mathcal M}_c \!\!&=&f_{\rho}m_{\rho}\!\! \int \frac{d^4k_{\text{\tiny $3$}}}{(2\pi)^4}g_{\text{\tiny $\Xi \Xi \rho$}}g_{\text{\tiny $P_{c} \Xi D$}} \frac{\bar u_{\text{\tiny ${P_c}$}}(p_{\text{\tiny $2$}})(/\!\!\!k_{\text{\tiny $3$}}+m_{\text{\tiny $\Xi$}})(\gamma_{\nu}-f_{\Xi\Xi\rho}\sigma_{\xi \nu }{k_{\text{\tiny $1$}}^{\xi}}/4m_{\rho})\nu_{\text{\tiny $\overline \Xi$}}(p_{\text{\tiny $1$}})}{-(k_{\text{\tiny $1$}}^2-m_{\text{\tiny $\rho$}}^2)\ (k_{\text{\tiny $2$}}^2-m_{\text{\tiny $D$}}^2)\ (k_{\text{\tiny $3$}}^2-m_{\Xi}^2)}\nonumber\\
&&(-g^{\nu \mu}+\frac{{k^{\nu}_1} {k^{\mu}_1}}{m_{\rho}^2}) \mathcal{F}^2(k_{\text{\tiny $3$}}^2)
 \Big((p+k_{\text{\tiny $2$}})_{\mu}F_1(k_{\text{\tiny $1$}}^2)\Big),\\
{\cal M}_{e }\!\!&=&if_{\pi}\!\!\int\frac{d^{4}k_{3}}{( 2\pi) ^{4}}g_{\text{\tiny $\Xi\Xi \pi$}}g_{\text{\tiny $P_{c}\Xi {D^*}$}}\frac{\bar u_{\text{\tiny $P_{c}$}}( p_{2})\gamma^{\beta}(/\!\!\!k_3-m_{\Xi} ){k\!\!\!/}_{1}\nu_{\text{\tiny $\overline \Xi$}}(p_{\text{\tiny $1$}})  (g_{\alpha\beta}-\frac{{{p_2}}_{\alpha}{{p_2}}_{\beta}}{m^{2}_{P_{c}}} ) (-g^{\alpha\mu}+\frac{k^{\mu}_{2}k^{\alpha}_{2}}{m^{2}_{D^*}})}{(k^{2}_{1}-m^{2}_{\pi} )(k^{2}_{2}-m^{2}_{D^{*}})(k^{2}_{3}-m^{2}_{\Xi} )  }\nonumber\\
&&\Big(-2m_{D^*}A_{1}(k^{2}_{1}){k_{1}}_{\mu}
+A_{3}( k^{2}_{1}) \frac{{ k_{1}}_{\mu} }{m_{B}+m_{D^*}}(p\cdot k_{1}+k_{2} \cdot k_{1}-m^{2}_{B}+m^{2}_{D^*})\Big)\mathcal{F}^2(k_{\text{\tiny $3$}}^2),\\
{\mathcal M}_{g}\!\!&=&f_{\rho}m_{\rho}\!\!\int\frac{d^{4}k_{3}}{2\pi ^{4}}g_{\text{\tiny $\Xi \Xi \rho$}}g_{\text{\tiny $P_{c}\Xi{D^*}$}}\frac{\bar u_{\text{\tiny $P_{c}$}}( p_{2}) \gamma_{5}\gamma^{\beta}(/\!\!\!k_3+m_{\Xi} )(\gamma_{\delta}-f_{\Xi\Xi\rho}\sigma_{\eta \delta}{k^{\eta}_1}/4m_{\rho}) \nu_{\text{\tiny $\overline \Xi$}}(p_{\text{\tiny $1$}}) }{-(k^{2}_{3}-m^{2}_{\Xi} )(k^{2}_{2}-m^{2}_{D^{*}} )(k^{2}_{1}-m^{2}_{\rho})}\nonumber\\&&
\Big(\frac{2iA_0(k^{2}_{1} ) }{m_{B}+m_{D^*}}\varepsilon_{\mu\rho\sigma\nu}{ k^{\rho}_{2}}p ^{\sigma}-A_{2}( k^{2}_{1}) g_{\nu\mu}
(m_{B}+m_{D^*})+A_{3}( k^{2}_{1})\frac{{ k_{1}}_{\nu} }{m_{B}+m_{D^*}}(p+k_{2} )_{\mu}\Big)\nonumber\\
&&\mathcal{F}^2(k_{\text{\tiny $3$}}^2)(g_{\alpha\beta}-\frac{{p_2}_{\alpha}{p_2}_{\beta}}{m^{2}_{P_{c}}} )(-g^{\alpha\nu}+\frac{{k^{\alpha}_2}{k^{\nu}_2}}{m_{D^*}})(-g^{\mu\delta}+\frac{{k^{\mu}_1}{k^{\delta}_1}}{m^{2}_{\rho}}).
\end{eqnarray}
The amplitude of $ B^-\to P_{c\bar udds}^-  \overline \Lambda^{0}$
\begin{eqnarray}
{\cal M}_a \!\!&=&i f_{\pi}\!\! \int \frac{d^4k_{\text{\tiny $3$}}}{(2\pi)^4}g_{\text{\tiny $\Lambda \Sigma \pi$}}g_{\text{\tiny $P_{c}\Sigma D  $}} \frac{\bar u_{\text{\tiny $P_{c}$}}(p_{\text{\tiny $2$}})(/\!\!\!k_{\text{\tiny $3$}}+m_{\text{\tiny $\Sigma$}})\gamma^{5}{k\!\!\!/}_{1}\nu_{\text{\tiny $\overline \Lambda$}}(p_{\text{\tiny $1$}})}{(k_{\text{\tiny $1$}}^2-m_{\text{\tiny $\pi$}}^2)\ (k_{\text{\tiny $2$}}^2-m_{\text{\tiny $D$}}^2)\ (k_{\text{\tiny $3$}}^2-m_{\text{\tiny $\Sigma$}}^2)}\nonumber\\&&\mathcal{F}^2(k_{\text{\tiny $1$}}^2)
 \Big(((2p_{\text{\tiny $2$}}+p_{\text{\tiny $1$}}-k_{\text{\tiny $3$}})\cdot (p_{\text{\tiny $1$}}+k_{\text{\tiny $3$}})-m^2_{\text{\tiny $B$}}+m^2_{\text{\tiny $D$}})F_1(k_{\text{\tiny $1$}}^2)+(m^2_{\text{\tiny $B$}}-m^2_{\text{\tiny $D$}})F_2(k_{\text{\tiny $1$}}^2)\Big),\\
{\mathcal M}_c \!\!&=&f_{\rho}m_{\rho}\!\! \int \frac{d^4k_{\text{\tiny $3$}}}{(2\pi)^4}g_{\text{\tiny $\Lambda \Sigma \rho$}}g_{\text{\tiny $P_{c}\Sigma D  $}} \frac{\bar u_{\text{\tiny ${P_c}$}}(p_{\text{\tiny $2$}})(/\!\!\!k_{\text{\tiny $3$}}+m_{\text{\tiny $\Sigma$}})(\gamma_{\nu}-f_{\Lambda\Sigma\rho}\sigma_{\xi \nu }{k_{\text{\tiny $1$}}^{\xi}}/4m_{\rho})\nu_{\text{\tiny $\overline \Lambda$}}(p_{\text{\tiny $1$}})}{-(k_{\text{\tiny $1$}}^2-m_{\text{\tiny $\rho$}}^2)\ (k_{\text{\tiny $2$}}^2-m_{\text{\tiny $D$}}^2)\ (k_{\text{\tiny $3$}}^2-m_{\Sigma}^2)}\nonumber\\
&&(-g^{\nu \mu}+\frac{{k^{\nu}_1} {k^{\mu}_1}}{m_{\rho}^2}) \mathcal{F}^2(k_{\text{\tiny $3$}}^2)
 \Big((p+k_{\text{\tiny $2$}})_{\mu}F_1(k_{\text{\tiny $1$}}^2)\Big),\\
{\cal M}_{e }\!\!&=&if_{\pi}\!\!\int\frac{d^{4}k_{3}}{( 2\pi) ^{4}}g_{\text{\tiny $\Lambda \Sigma \pi$}}g_{\text{\tiny $P_{c}\Sigma D^* $}} \frac{\bar u_{\text{\tiny $P_{c}$}}( p_{2})\gamma^{\beta}(/\!\!\!k_3-m_{\Sigma} ){k\!\!\!/}_{1}\nu_{\text{\tiny $\overline \Lambda$}}(p_{\text{\tiny $1$}})  (g_{\alpha\beta}-\frac{{{p_2}}_{\alpha}{{p_2}}_{\beta}}{m^{2}_{P_{c}}} ) (-g^{\alpha\mu}+\frac{k^{\mu}_{2}k^{\alpha}_{2}}{m^{2}_{D^*}})}{(k^{2}_{1}-m^{2}_{\pi} )(k^{2}_{2}-m^{2}_{D^{*}})(k^{2}_{3}-m^{2}_{\Sigma} )  }\nonumber\\
&&\Big(-2m_{D^*}A_{1}(k^{2}_{1}){k_{1}}_{\mu}
+A_{3}( k^{2}_{1}) \frac{{ k_{1}}_{\mu} }{m_{B}+m_{D^*}}(p\cdot k_{1}+k_{2} \cdot k_{1}-m^{2}_{B}+m^{2}_{D^*})\Big)\mathcal{F}^2(k_{\text{\tiny $3$}}^2),\\
{\mathcal M}_{g}\!\!&=&f_{\rho}m_{\rho}\!\!\int\frac{d^{4}k_{3}}{2\pi ^{4}}g_{\text{\tiny $\Lambda \Sigma \rho$}}g_{\text{\tiny $P_{c}\Sigma D^* $}} \frac{\bar u_{\text{\tiny $P_{c}$}}( p_{2}) \gamma_{5}\gamma^{\beta}(/\!\!\!k_3+m_{\Sigma} )(\gamma_{\delta}-f_{\Lambda\Sigma\rho}\sigma_{\eta \delta}k^{\eta}_{1}/4m_{\rho}) \nu_{\text{\tiny $\overline \Lambda$}}(p_{\text{\tiny $1$}}) }{-(k^{2}_{3}-m^{2}_{\Sigma} )(k^{2}_{2}-m^{2}_{D^{*}} )(k^{2}_{1}-m^{2}_{\rho})}\nonumber\\&&
\Big(\frac{2iA_0(k^{2}_{1} ) }{m_{B}+m_{D^*}}\varepsilon_{\mu\sigma\rho \nu}{ k^{\rho}_{2}}p ^{\sigma}-A_{2}( k^{2}_{1}) g_{\nu\mu}
(m_{B}+m_{D^*})+A_{3}( k^{2}_{1})\frac{{ k_{1}}_{\nu} }{m_{B}+m_{D^*}}(p+k_{2} )_{\mu}\Big)\nonumber\\
&&\mathcal{F}^2(k_{\text{\tiny $3$}}^2)(g_{\alpha\beta}-\frac{{p_2}_{\alpha}{p_2}_{\beta}}{m^{2}_{P_{c}}} )(-g^{\alpha\nu}+\frac{{k^{\alpha}_2}{k^{\nu}_2}}{m_{D^*}})(-g^{\mu\delta}+\frac{{k^{\mu}_1}{k^{\delta}_1}}{m^{2}_{\rho}}).
\end{eqnarray}
The amplitude of $\overline B^0\to P_{c\bar duss}^+ \overline \Sigma^{-}$
\begin{eqnarray}
{\cal M}_a \!\!&=&i f_{\pi}\!\! \int \frac{d^4k_{\text{\tiny $3$}}}{(2\pi)^4}g_{\text{\tiny $\Xi \Sigma K$}}g_{\text{\tiny $P_{c}\Xi D $}} \frac{\bar u_{\text{\tiny $P_{c}$}}(p_{\text{\tiny $2$}})(/\!\!\!k_{\text{\tiny $3$}}+m_{\text{\tiny $\Xi$}})\gamma^{5}{k\!\!\!/}_{1}\nu_{\text{\tiny $\overline \Sigma$}}(p_{\text{\tiny $1$}})}{(k_{\text{\tiny $1$}}^2-m_{\text{\tiny $K$}}^2)\ (k_{\text{\tiny $2$}}^2-m_{\text{\tiny $D$}}^2)\ (k_{\text{\tiny $3$}}^2-m_{\text{\tiny $\Xi$}}^2)}\nonumber\\&&\mathcal{F}^2(k_{\text{\tiny $1$}}^2)
 \Big(((2p_{\text{\tiny $2$}}+p_{\text{\tiny $1$}}-k_{\text{\tiny $3$}})\cdot (p_{\text{\tiny $1$}}+k_{\text{\tiny $3$}})-m^2_{\text{\tiny $B$}}+m^2_{\text{\tiny $D$}})F_1(k_{\text{\tiny $1$}}^2)+(m^2_{\text{\tiny $B$}}-m^2_{\text{\tiny $D$}})F_2(k_{\text{\tiny $1$}}^2)\Big),\\
{\mathcal M}_c \!\!&=&f_{\rho}m_{\rho}\!\! \int \frac{d^4k_{\text{\tiny $3$}}}{(2\pi)^4}g_{\text{\tiny $\Xi \Sigma K^*$}}g_{\text{\tiny $P_{c}\Xi D $}} \frac{\bar u_{\text{\tiny ${P_c}$}}(p_{\text{\tiny $2$}})(/\!\!\!k_{\text{\tiny $3$}}+m_{\text{\tiny $\Xi$}})(\gamma_{\nu}-f_{\Xi\Sigma K^*}\sigma_{\xi \nu }{k_{\text{\tiny $1$}}^{\xi}}/4m_{\rho})\nu_{\text{\tiny $\overline \Sigma$}}(p_{\text{\tiny $1$}})}{-(k_{\text{\tiny $1$}}^2-m_{\text{\tiny $K^*$}}^2)\ (k_{\text{\tiny $2$}}^2-m_{\text{\tiny $D$}}^2)\ (k_{\text{\tiny $3$}}^2-m_{\Xi}^2)}\nonumber\\
&&(-g^{\nu \mu}+\frac{{k^{\nu}_1} {k^{\mu}_1}}{m_{K^*}^2}) \mathcal{F}^2(k_{\text{\tiny $3$}}^2)
 \Big((p+k_{\text{\tiny $2$}})_{\mu}F_1(k_{\text{\tiny $1$}}^2)\Big),\\
{\cal M}_{e }\!\!&=&if_{\pi}\!\!\int\frac{d^{4}k_{3}}{( 2\pi) ^{4}}g_{\text{\tiny $\Xi \Sigma K$}}g_{\text{\tiny $P_{c}\Xi D^*$}} \frac{\bar u_{\text{\tiny $P_{c}$}}( p_{2})\gamma^{\beta}(/\!\!\!k_3-m_{\Xi} ){k\!\!\!/}_{1}\nu_{\text{\tiny $\overline \Sigma$}}(p_{\text{\tiny $1$}})  (g_{\alpha\beta}-\frac{{{p_2}}_{\alpha}{{p_2}}_{\beta}}{m^{2}_{P_{c}}} ) (-g^{\alpha\mu}+\frac{k^{\mu}_{2}k^{\alpha}_{2}}{m^{2}_{D^*}})}{(k^{2}_{1}-m^{2}_{K} )(k^{2}_{2}-m^{2}_{D^{*}})(k^{2}_{3}-m^{2}_{\Xi} )  }\nonumber\\
&&\Big(-2m_{D^*}A_{1}(k^{2}_{1}){k_{1}}_{\mu}
+A_{3}( k^{2}_{1}) \frac{{ k_{1}}_{\mu} }{m_{B}+m_{D^*}}(p\cdot k_{1}+k_{2} \cdot k_{1}-m^{2}_{B}+m^{2}_{D^*})\Big)\mathcal{F}^2(k_{\text{\tiny $3$}}^2),\\
{\mathcal M}_{g}\!\!&=&f_{\rho}m_{\rho}\!\!\int\frac{d^{4}k_{3}}{2\pi ^{4}}g_{\text{\tiny $\Xi \Sigma K^*$}}g_{\text{\tiny $P_{c}\Xi D^* $}} \frac{\bar u_{\text{\tiny $P_{c}$}}( p_{2}) \gamma_{5}\gamma^{\beta}(/\!\!\!k_3+m_{\Xi} )(\gamma_{\delta}-f_{\Xi\Sigma K^*}\sigma_{\eta \delta}k^{\eta}_{1}/4m_{\rho}) \nu_{\text{\tiny $\overline \Sigma$}}(p_{\text{\tiny $1$}}) }{-(k^{2}_{3}-m^{2}_{\Xi} )(k^{2}_{2}-m^{2}_{D^{*}} )(k^{2}_{1}-m^{2}_{K^*})}\nonumber\\&&
\Big(\frac{2iA_0(k^{2}_{1} ) }{m_{B}+m_{D^*}}\varepsilon_{\mu\rho\sigma \nu}{ k^{\rho}_{2}}p ^{\sigma}-A_{2}( k^{2}_{1}) g_{\nu\mu}
(m_{B}+m_{D^*})+A_{3}( k^{2}_{1})\frac{{ k_{1}}_{\nu} }{m_{B}+m_{D^*}}(p+k_{2} )_{\mu}\Big)\nonumber\\
&&\mathcal{F}^2(k_{\text{\tiny $3$}}^2)(g_{\alpha\beta}-\frac{{p_2}_{\alpha}{p_2}_{\beta}}{m^{2}_{P_{c}}} )(-g^{\alpha\nu}+\frac{{k^{\alpha}_2}{k^{\nu}_2}}{m_{D^*}})(-g^{\mu\delta}+\frac{{k^{\mu}_1}{k^{\delta}_1}}{m^{2}_{K^*}}).
\end{eqnarray}
The amplitude of $\overline B_s^0\to P_{c\bar suud}^{++}  \overline \Sigma^{--}_{\bar c}$
\begin{eqnarray}
{\cal M}_b \!\!&=&(i f_{D_s})\!\! \int \frac{d^4k_{\text{\tiny $3$}}}{(2\pi)^4}g_{\text{\tiny $\Sigma_c N D$}}g_{\text{\tiny $P_{c}ND_s$}} \frac{\bar u_{\text{\tiny $P_{c}$}}(p_{\text{\tiny $2$}})(/\!\!\!k_{\text{\tiny $3$}}+m_{\text{\tiny $N$}})\gamma^{5}{k\!\!\!/}_{1}\nu_{\text{\tiny $\overline \Sigma_{\bar c}$}}(p_{\text{\tiny $1$}})}{(k_{\text{\tiny $1$}}^2-m_{\text{\tiny $D_s$}}^2)\ (k_{\text{\tiny $2$}}^2-m_{\text{\tiny $D$}}^2)\ (k_{\text{\tiny $3$}}^2-m_{\text{\tiny $N$}}^2)}\nonumber\\&&\mathcal{F}^2(k_{\text{\tiny $3$}}^2)
 (2p_{\text{\tiny $2$}}+p_{\text{\tiny $1$}}-k_{\text{\tiny $3$}})\cdot (p_{\text{\tiny $1$}}+k_{\text{\tiny $3$}})-m^2_{\text{\tiny $B$}}+m^2_{\text{\tiny $D$}})F_1(k_{\text{\tiny $1$}}^2)+(m^2_{\text{\tiny $B$}}-m^2_{\text{\tiny $D$}})F_2(k_{\text{\tiny $1$}}^2)\Big),\\
{\cal M}_d \!\!&=&(i f_{D_s})\!\! \int \frac{d^4k_{\text{\tiny $3$}}}{(2\pi)^4}g_{\text{\tiny $\Sigma_c N D^*$}}g_{\text{\tiny $P_{c} ND_s $}} \frac{\bar u_{\text{\tiny $P_{c}$}}(p_{\text{\tiny $2$}})(/\!\!\!k_{\text{\tiny $3$}}+m_{\text{\tiny $N$}})(\gamma_{\nu}-f_{\Sigma_c N D^*}\sigma_{m \nu}k_{\text{\tiny $1$}}^{m}/4m_{D^*})\nu_{\text{\tiny $\overline \Sigma_{\bar c}$}}(p_{\text{\tiny $1$}})(-g^{\mu \nu}+\frac{{{k^{\mu}_2}} {{k^{\nu}_2}}}{m_{D^*}^2})}{(k_{\text{\tiny $1$}}^2-m_{\text{\tiny $D_s$}}^2)\ (k_{\text{\tiny $2$}}^2-m_{\text{\tiny $D^*$}}^2)\ (k_{\text{\tiny $3$}}^2-m_{N}^2)}\nonumber\\
&&\mathcal{F}^2(k_{\text{\tiny $3$}}^2) \Big(-2m_{D^*}A_{1}(k^{2}_{1}){{k_1}_{\mu}}
+A_{2}( k^{2}_{1}) \frac{{k_1}_{\mu} }{m_{B}+m_{D^*}}(p\cdot k_{1}+k_{2} \cdot k_{1}-m^{2}_{B}+m^{2}_{D^*})\Big).
\end{eqnarray}
Where $m_{\Xi}$, $m_{\Sigma}$, $m_{N}$ and $k_{3}$ represent the mass and momentum of exchanged baryon($\Xi$,$\Sigma$,$N$) respectively.
 
 \end{document}